%%%%%%%%%%%%%%%%%%%%%%%%%%%%%%%%%%%%%%%%%%%%%%%%%%%%%%%%%%%%%%%%%%%%%%%%%%%
\input harvmac
\noblackbox
\def\ajou#1&#2(#3){\ \sl#1\bf#2\rm(19#3)}
\def\jou#1&#2(#3){\unskip, \sl#1\bf#2\rm(19#3)}
\def\gtwid{\mathrel{\raise.3ex\hbox{$>$\kern-.75em\lower1ex\hbox{$\sim$}}}}
\def\ltwid{\mathrel{\raise.3ex\hbox{$<$\kern-.75em\lower1ex\hbox{$\sim$}}}}
\def\Title#1#2{\rightline{#1}\ifx\answ\bigans\pageno1\vskip1in
\else\pageno1\vskip.8in\fi \centerline{\titlefont #2}\vskip .5in}

scaled\magstep3
 
scaled\magstep3
\font\ticp=cmcsc10
\def\frac#1#2{{#1 \over #2}}
\def\sq{{\vbox {\hrule height 0.6pt\hbox{\vrule width 0.6pt\hskip 3pt
   \vbox{\vskip 6pt}\hskip 3pt \vrule width 0.6pt}\hrule height 0.6pt}}}
\def\s{\sigma}

\def\[{\left [}
\def\]{\right ]}
\def\({\left (}
\def\){\right )}

\def\p{\partial}
\def\E{E}

%references
\lref\mtw{C.~Misner, K.~Thorne, and J.~Wheeler, {\it Gravitation},
Freeman \& Wiley (1972).}
\lref\fut{A useful book with references to the extensive original
literature is J.F.~Futterman, F.~Handler and R.~Matzner, {\it Scattering
from Black Holes}, Cambridge University Press (1988).}
\lref\bmpv{J.~Breckenridge, R.~Myers, A.~Peet, and C.~Vafa, {\it
$D$-Branes and Spinning Black Holes};
hep-th/9602065.}
\lref\hml{G.~Horowitz, D.~Lowe, and J.~Maldacena, {\it Statistical
Entropy of Nonextremal Four-Dimensional Black Holes and $U$-Duality}
 Phys.~Rev~Lett. {\bf 77}, (1996) 430; hep-th/9603195.} 
\lref\gihu{G.~Gibbons and C.~Hull, {\it 
A Bogomolny Bound for General Relativiy and Solitons in N=2 
Supergravity}, {\sl Phys~Lett.} {\bf B109}, (1982), 190.}

\lref\pgf{D.~Page, \jou Phys~Rev. & D14 (76) 1509.}

\lref\dasmathurtwo{S. Das and S. Mathur,
``{\it Interactions Involving D-branes}'',
hep-th/9607149.}

\lref\gk{ S. Gubser and I. Klebanov, 
{\it Emission of Charged
Particles
from Four- and Five-dimensional Black Holes}, 
hep-th/9608108; {\it Four-dimensional greybody 
factors and the effective string},  hep-th/9609076. }

\lref\page{ D. Page, Phys. Rev. {\bf D 13} (1976) 198; Phys. Rev. {\bf
D14}
(1976), 3260.}

\lref\unruh{W. Unruh, 
Phys. Rev. {\bf D14} (1976) 3251.}
\lref\cj{ M. Cvetic and D. Youm, hep-th/9512127.}

\lref\tseytlin{A. Tseytlin, hep-th/9601119.}
\lref\tata{S. Dhar, G. Mandal and S. Wadia, ``{\it
Absorption vs Decay of Black Holes in String Theory and 
T-symmetry}'', hep-th/9605234.}
\lref\gm{G. Horowitz and D. Marolf, hep-th/9605224.}
\lref\mss{J. Maldacena and L. Susskind, {\it D-branes and Fat 
Black Holes }, Nucl. Phys. {\bf B475} (1996) 679,  hep-th/9604042.}
\lref\bl{V. Balasubramanian and F. Larsen, hep-th/9604189.}
\lref\jmn{J. Maldacena,``{\it Statistical Entropy of Near Extremal
Fivebranes}'', hep-th/9605016.}
\lref\hlm{G. Horowitz, D. Lowe and J. Maldacena, hep-th/9603195.}
\lref\ms{J. Maldacena and A. Strominger, hep-th/9603060.}
\lref\bd{See N. D. Birrel and P.C. Davies,``{\it Quantum Fields in
Curved
Space}'', Cambridge University Press 1982.}
\lref\hms{G. Horowitz, J. Maldacena and A. Strominger,
``{\it Nonextremal Black Hole Microstates and U-duality}'',
 hep-th/9603109.}
\lref\dbr{J. Polchinski, S. Chaudhuri, and C. Johnson, hep-th/9602052.}
\lref\jp{J. Polchinski, hep-th/9510017.}
\lref\dm{S. Das and S. Mathur, {\it
Comparing Decay Rates for Black Holes and D-branes},hep-th/9606185.}
\lref\ghas{G. Horowitz and A. Strominger, {\it Counting States of
Near-extremal Black Holes}, Phys. Rev. Lett. {\bf 77} (1996) 2368,
  hep-th/9602051.}
\lref\ascv{A. Strominger and C. Vafa, {\it On the Microscopic 
Origin of the Bekenstein-Hawking Entropy}, Phys. Lett. {\bf B379} (1996)
99,
 hep-th/9601029.}
\lref\hrva{P. Horava, Phys. Lett. {\bf B231} (1989) 251.}
\lref\cakl{C. Callan and I. Klebanov, hep-th/9511173.}
\lref\prskll{J. Preskill, P. Schwarz, A. Shapere, S. Trivedi and
F. Wilczek, Mod. Phys. Lett. {\bf A6} (1991) 2353. }
\lref\bhole{G. Horowitz and A. Strominger,
Nucl. Phys. {\bf B360} (1991) 197.}
\lref\bekb{J. Bekenstein, Phys. Rev {\bf D12} (1975) 3077.}
\lref\hawkirr{S. Hawking, Phys. Rev {\bf D13} (1976) 191.}
\lref\stas{A.~Strominger and S.~Trivedi,  Phys.~Rev. {\bf D48}
 (1993) 5778.}
\lref\bek{J. Bekenstein, Lett. Nuov. Cimento {\bf 4} (1972) 737,
Phys. Rev. {\bf D7} (1973) 2333, Phys. Rev. {\bf D9} (1974) 3292.}
\lref\hawkb{S. Hawking, Nature {\bf 248} (1974) 30, Comm. Math. Phys.
{\bf 43} (1975) 199.}
\lref\cm{C. Callan and J. Maldacena, {\it The D-brane approach to 
black hole quantum mechanics}, Nucl. Phys. {\bf B 475 } (1996)
645, hep-th/9602043.}
\lref\ka{ B. Kol and A. Rajaraman, ``{\it
Fixed Scalars and Suppression of Hawking Evaporation}'',  hep-th/9608126. }
\lref\vafainst{ M. Bershadsky, V. Sadov and  C. Vafa,  hep-th/9511222;
C.  Vafa, hep-th/9512078.}
\lref\becspin{  J. C. Breckenridge, D. A. Lowe, 
R. C. Myers, A. W. Peet, A. Strominger and  C. Vafa, ``{\it
Macroscopic and Microscopic Entropy of Near-Extremal Spinning Black
Holes}'',  hep-th/9603078.} 
\lref\jmas{J. Maldacena and A. Strominger, {\it Black Hole 
Greybody Factors and D-brane Spectroscopy},  hep-th/9609026.}
\lref\cgkt{C. Callan, S. Gubser, I. Klebanov and A. Tseytlin,
{\it Absorption
of Fixed Scalars and the D-Brane Approach to Black Holes},
hep-th/9610172.}

\lref\gibbons{S. Das, G. Gibbons and S. Mathur, {\it 
Universality of Low Energy Absorption Cross Sections for 
Black Holes}, hep-th/9609052.}

\lref\jmmlow{J. Maldacena, {\it D-branes and Near Extremal Black Holes},
hep-th/9611125.}
\lref\lenny{L. Susskind, {\it Some Speculations about 
Black Hole Entropy in String Theory}, hep-th/9309145.
}
\lref\ssc{E. Halyo, B. Kol, A. Rajaraman and L. Susskind, 
{\it Counting Schwarzschild and Charged Black Holes}, hep-th/9609075.}

\lref\steve{S. Gubser, unpublished notes. }

% \draft

\Title{\vbox{\baselineskip12pt\hbox{UCSBTH-97-02}
\hbox{RU-97-03}\hbox{hep-th/9702015}
% \hbox{Draft -- do not distribute}
}}
{\vbox{\centerline {Universal Low-Energy Dynamics}
\vskip .1in
\centerline {for Rotating Black Holes}
}}
\centerline{{\ticp Juan Maldacena}\footnote{$^*$}
{Email address:
malda@physics.rutgers.edu} and 
{\ticp Andrew Strominger}\footnote{$^\dagger$}
{Email address:
andy@denali.physics.ucsb.edu} }
\vskip.1in
\centerline{\sl $^*$Serin Physics Labs, Rutgers University,
 Piscataway, NJ 08855-0849}
\centerline{\sl $^\dagger$Department of Physics, University of
California, Santa Barbara, CA 93106-9530}

\bigskip
\centerline{\bf Abstract}

Fundamental string theory has been used to show that low energy
excitations of certain black holes are described by a
two dimensional conformal field theory.
This picture has been found to be extremely
robust. 
In this paper it is argued that many essential features of the 
low energy effective theory can be inferred directly 
from a semiclassical analysis of the general Kerr-Newman solution
of supersymmetric four-dimensional Einstein-Maxwell gravity, 
without using string theory.
We consider the absorption 
and emission  of scalars with orbital angular momentum, which provide  
a sensitive probe of the black hole.  We find that the semiclassical 
emission rates -including superradiant emission and greybody factors -
for such scalars agree in striking detail with those computed in the 
effective conformal field theory,  in both four and five dimensions. 
Also the value of the  quantum mass gap to 
the lowest-lying
excitation of a charge-$Q$ black hole, 
$E_{gap}=1/8Q^3$ in Planck units, can be derived without knowledge
of fundamental string theory.

% It has a range of validity far exceeding the range of validity
%of its original derivation from fundamental string theory.  

\vfill
\eject

{\baselineskip=24pt
\newsec{Introduction}

Recent statistical derivations of the Bekenstein-Hawking entropy have
used weakly-coupled fundamental string theory as a starting point
\ascv \cm \ghas  \hml . In
full detail, the derivation is not simple and requires a precise
understanding of string theory and $D$-branes. The final answer is,
however, much simpler than the derivation: The quantum states of a
near-BPS black hole are  described by a low energy  supersymmetric
conformal field theory or effective string whose parameters are 
functions of the charges.
 Furthermore, the validity of this effective string picture
 extends  far beyond the domain of validity of its original derivation
from fundamental string theory.  Indeed it gives accurate
decay
rates in the $M$-theory region where there are no fundamental strings at
all!

How did this happen? On general grounds, one expects the near-BPS
dynamics of a black hole to be described by some effective field theory,
whether or not string theory is weakly coupled.  Apparently we have
stumbled upon the effective black hole field theory which is valid,
at sufficiently low energies,  for
all values of the string coupling.

Given this state of affairs, it is natural to ask: how much of 
this effective
theory could have been discovered {\it without} knowledge of fundamental
string theory?  In this paper we address this question in the simple
context of four-dimensional, Einstein-Maxwell gravity.  We begin by
assuming that, on scales large compared to the Schwarzschild radius, there
is some kind of weakly-coupled, unitary effective field theory. We then
demand consistency of this effective theory with
semi-classical, black-hole thermodynamics and decay rates.  This is a
highly over-constrained problem, especially when decay rates into 
channels with non-zero angular momentum are considered. 
The effective superstring theory
provides  a solution, possibly the only one.
 Hence we conclude that --- had
history been a little different --- much of 
the effective string picture of
black-hole dynamics might have been derived without knowledge of
fundamental string theory. Of course for a complete and 
systematic picture string 
theory remains essential. 
%This approach shows that the low energy dynamics of a black hole
%can be described in terms of a unitary field theory.

The four dimensional case is considered in section 2. 
In sections 2.1-2.10 we semiclassically compute 
the absorption cross section 
and decay rates for a massless scalar with angular momentum 
and a near-BPS Kerr-Newman black hole. This depends on five 
parameters: the 
mass, charge and angular momentum of the black hole, as well as the 
frequency and  angular momentum of the scalar. The total emission includes 
superradiant emission, which occurs for a rotating black hole even
at extremality when the Hawking temperature vanishes. In 2.11 
we argue that the Kerr-Newman entropy formula - including rotation -
implies that the black hole degrees of freedom relevant for near-BPS 
excitations can be described by 
a $(0,4)$ chiral superconformal field theory with an
SU(2) current algebra associated to rotations. 
We determine the level
of the current algebra by equating the bound
on $L_0$ in terms of the SU(2) charge with the bound implied
by the absence of a naked singularity.
The mass gap is then
computed as the
energy of the first excited state of this theory. In section 2.12
the proposed conformal field theory is used to compute the decay
rates.
It turns out that the decay rates are almost completely determined
by general properties of the two dimensional field theory correlators.
Comparison  with the semiclassical results of sections 2.1-2.9  
reveals detailed agreement. 

The five dimensional case is considered in section 3. 
An important new feature
here is that angular momentum can be carried by both left and right 
movers on the effective string. In particular an $\ell=1$ boson 
can be emitted by the collision of left and right 
moving $\ell=1/2$ fermions. The rate for this in the effective string 
picture involves a right and a left $\it fermionic$ thermal 
occupation factors. In the semiclassical picture such  factors could
come only from the greybody effects. We will see that such 
factors indeed arise with exactly the right form. Hence one 
can directly `see' the fermionic constituents of a black hole 
in the $\ell=1$ scalar emission spectrum!

\newsec{The Four-Dimensional Kerr-Newman Black Hole}

\subsec{The Classical Geometry}

The metric for a black hole of charge $Q$, mass $M$, and angular
momentum $J=Ma$ is 
\eqn\mtw{\eqalign{ds^2 & = 
- \left(\frac{\Delta - a^2 \sin^2\theta}{\Sigma}\right)
dt^2 - \left(\frac{2 a \sin^2 \theta (r^2+ a^2 -\Delta)}{\Sigma}\right)
dt d \phi\cr
&+ \left(\frac{(r^2+a^2)^2 - \Delta a^2 \sin^2 \theta}{\Sigma}\right)
\sin^2 \theta d \phi^2\cr
& + \frac{\Sigma}{\Delta}\ dr^2 + \Sigma\ d\, \theta^2\ ,}}
where
\eqn\ddef{\eqalign{\Sigma & \equiv r^2 + a^2 \cos^2 \theta\ ,\cr
\Delta & \equiv r^2 + a^2 + Q^2 - 2Mr\ .}}
and we are setting the Planck length to one and  $G_N=1$.
%Other useful metric quantities are
%
%\eqn\twothree{\eqalign{\sqrt{-g} = \sin\, \theta\Sigma\quad &, \quad 
%g^{tt} = \frac{\Delta a^2 \sin^2 \theta -r^2 -a^2}{\Delta\Sigma}\ ,\cr
%g^{t\phi} = \frac{a(\Delta - r^2 - a^2)}{\Delta\Sigma} \quad &, \quad 
%g^{\phi\phi} = \frac{\Delta - a^2 \sin^2 \theta}{\Delta \Sigma \sin^2
%\theta}\ .}}
%
The inner and outer horizons are located at the zeroes of $\Delta$:
\eqn\twofour{r_\pm = M \pm \sqrt{M^2 - Q^2 - a^2}\ .}
The area ${\cal A}$, Hawking temperature $T_H$, angular velocity $\Omega$
and electric potential $\Phi$ at the horizon are
\eqn\thdef{\eqalign{{\cal A} & = 4\pi \left(2M^2 - Q^2 + 2M 
\sqrt{M^2-Q^2-a^2}\right)=4\pi r_+^2
\ ,\cr
T_H & = \frac{ (r_+ - r_-)}{{\cal A}}\ ,\cr
\Omega & = \frac{4\pi a}{{\cal A}}\ , \cr
\Phi & = \frac{4\pi Q r_+}{{\cal A}}\ .}}
These quantities are related by the first law
\eqn\twosix{dM = T_H dS + \Omega d J + \Phi d Q\ .}
where the entropy is $S= {\cal A}/4$.
\subsec{The Scalar Wave Equation}

In this section we give the separated form of the wave equation $\sq\Phi=0$
for a massless scalar. As for the well-studied case of Kerr \refs{\fut},
the solution separates as
\eqn\threeone{\Phi= e^{im\phi-iwt} \ S^m_A (\theta; a\omega) R (r)\ .}
$S$ obeys
\eqn\sae{\left(\frac{1}{\sin\theta}\ \partial_\theta \sin \theta
\partial_\theta  - \frac{m^2}{\sin^2\theta} + a^2\omega^2 \cos^2\theta\right)
\ S^m_A (\theta; a\omega) = - A\ S^m_A (\theta; a\omega)}
For small $aw$ (the case of eventual interest to us) the eigenvalues
are
\eqn\threethree{A= \ell (\ell+1) + {\cal O}(a^2w^2)\ .}
$R$ then obeys
\eqn\weq{\Delta \partial_r \Delta \partial_r R + K^2 R - \lambda \Delta
R=0\ ,}
where
\eqn\hldef{\eqalign{K & \equiv \omega(r^2 + a^2) - ma\ ,\cr
\lambda & \equiv A + a^2w^2 - 2 m\omega a\ .}}

\subsec{ Low-Frequency Scalar Absorption}

In this section will calculate the low energy absorption 
cross section for the black holes described in section 2.1.
The low energy condition is $\omega \ll 1/M$, which means that the
Compton wavelength of the particle is much bigger than the
gravitational size of the black hole, defined as the place where
the redshift between a static observer and an the asymptotic observer
becomes of order 1. 
We also assume that $\Omega \ll 1/M $ for simplicity.

We use a matching procedure,
  dividing  the spacetime outside the horizon,
 $r_+ \le r $ into two overlapping regions defined by 

\item{} Near Region: $r-r_+ \ll 1/\omega$,
\item{} Far Region: $M \ll r-r_+$.

In each  region the wave equation can be approximated using the
inequalities and then exactly solved. A complete solution can then be
obtained by matching. 
We now discuss each region in turn.

\subsec{Near Region Wave Equation}

In the near region, the coordinate distance $r-r_+$  is small compared
with the inverse frequency $1/\omega$.
This implies that we can replace the functions $K^2 - \lambda \Delta $
in
\weq\ by 
\eqn\fiveone{K^2 - \lambda \Delta \approx
 r_+^4 (\omega-m\Omega)^2 -\ell (\ell + 1)\Delta}
where the angular velocity $\Omega$ of the black hole is given  in
\thdef .  We have approximated $K$ by its
constant value at small $r \sim r_+$ since  the $r$ dependence of the
potential
is dominated by the term proportional to $\Delta$, and we have also
neglected 
the  term involving $\omega a^2 $ in  \hldef .   
We can  approximate the 
eigenvalues of the angular Laplacian in \sae\ and 
$\lambda$ in \hldef\ by $\ell (\ell +1)$.
The equation \weq\ is then approximately
\eqn\neq{\Delta \partial_r \Delta \partial_r R + r_+^4 (\omega -
m\Omega)^2 R-\ell (\ell+1) \Delta R=0\ .}

\subsec{Far Region Wave Equation}

In this region we are far from the black hole and its effects disappear.
One has simply
\eqn\feq{\frac{1}{r^2} \partial_r r^2 \partial_r R + \omega^2 R -
\frac{\ell(\ell+1)}{r^2} R=0\ ,}
the equation for a massless scalar field of frequency $\omega$ and
angular momentum $\ell$ in flat spacetime.

\subsec{Near Region Solution}

In order to solve the near region equation, we define a new variable
\eqn\fivesix{z= \frac{r-r_+}{r-r_-}\quad , \quad 0\leq z \leq 1\ .}
The horizon is at $z=0$. One finds 
\eqn\fiveseven{\Delta \partial_r = (r_+ - r_-) z \partial_z\ .}
The near-region wave equation \neq\ is then
\eqn\fiveeight{z (1-z) \partial^2_z R + (1-z) \partial_z R +
\left(\frac{\omega-m\Omega}{4\pi T_H}\right)^2 \left(1 + \frac{1}{z}\right)
R - \frac{\ell(\ell+1)}{1-z}\ R=0 \ .}
this can be transformed into the standard hypergeometric form by defining
\eqn\hfn{R=Az^{i\frac{\omega-m\Omega}{4\pi T_H}} (1-z)^{\ell+1} F\ ,}
where $A$ is a to-be-determined normalization constant. $F$ then obeys
\eqn\fiveten{\eqalign{z(1-z) \partial^2_z F &+ \left(1+i
\frac{\omega-m\Omega}{2\pi T_H} - (1+2(\ell+1) + i
\frac{\omega-m\Omega}{2\pi T_H})z \right) \partial_z F\cr
& - \left((\ell + 1)^2 + i
\ \frac{\omega-m\Omega}{2\pi T_H}\ (\ell + 1)\right) F=0\ .}}
Since we are interested in calculating the absorption cross section
we impose the condition that there is only ingoing 
flux at the horizon $z=0$. This implies that $F$ in \hfn\  
 is the standard hypergeometric function
$F(\alpha,\beta,\gamma;z)$ with
\eqn\fivetwelve{\eqalign{\alpha & =
\ell+1+i\frac{\omega-m\Omega}{2\pi T_H}\ ,\cr
\beta & = \ell + 1\ ,\cr
\gamma & = 1+i\frac{\omega-m\Omega}{2\pi T_H}\ .}}

\subsec{Far Region Solution}

The far region solution is a linear combination of Bessel functions
\eqn\bfn{R = \frac{1}{\sqrt{r}} \left[\alpha J_{\ell + \half}
(\omega r) + \beta J_{-\ell-\half} (\omega r)\right] \ .}
For large $R$ this behaves as
\eqn\bas{R \buildrel{r\to\infty}\over\to  \frac{1}{r}\
\sqrt{\frac{2}{\pi\omega}} \left[-\alpha \sin\ \left(\omega r -
\frac{\ell\pi}{2}\right) + \beta \cos
\ \left(\omega r + \frac{\ell\pi}{2}\right)\right] \ .}

\subsec{Matching the Far and Near Solutions}

Next we need to match the small $r$ far region Bessel functions
 \bfn\ to
the large $r$ $(z\to 1)$ near region hypergeometric function.  At
small $r$ \bfn\ behaves as
\eqn\smr{R \simeq \frac{1}{\sqrt{r}} 
\left[\alpha \left(\frac{\omega r}{2}\right)^{\ell
+\half} \frac{1}{\Gamma(\ell + \frac{3}{2})} + \beta \left(\frac{\omega
r}{2}\right)^{-\ell-\half} \frac{1}{\Gamma(-\ell + \half)}\right]\ .}
with corrections to both terms suppressed by $r^2$. The large $r$,
$z\to1$ behavior of the near region solution \hfn\ follows from
the $z\to1-z$ transformation law for hypergeometric functions
\eqn\hgf{\eqalign{F(\alpha, \beta; \gamma; z) & =
\frac{\Gamma(\gamma)\Gamma(\gamma-\alpha-\beta)}{\Gamma(\gamma-\alpha)\Gamma
(\gamma-\beta)}\ F(\alpha, \beta; \alpha + \beta - \gamma + 1; 1-z)\cr
& + (1-z)^{\gamma-\alpha-\beta}
\ \frac{\Gamma(\gamma)\Gamma(\alpha + \beta -
\gamma)}{\Gamma(\alpha)\Gamma(\beta)}\ F(\gamma-\alpha, \gamma-\beta;
 \gamma-\alpha - \beta +1; 1-z)\ .}}
Using $1-z\to (r_+ - r_-)/r$, one finds that for large $r$ \hfn\ is given
by
\eqn\lgr{\eqalign{R= A\left(\frac{r}{r_+-r_-}\right)^{-\ell-1}& \Gamma
\left(1+i\frac{\omega-m\Omega}{2\pi T_H}\right) \times \cr
\biggl(\frac{\Gamma(-2\ell-1)}{\Gamma (-\ell)\Gamma\left(-\ell+i\frac{\omega
-m\Omega}{2\pi T_H}\right)} & + \left(\frac{r}{r_+ - r_-}\right)^{2\ell+1}
\frac{\Gamma(2\ell+1)}{\Gamma(\ell+1)\Gamma
\left(\ell+1+i\frac{\omega-m\Omega}{2\pi T_H}\right)}
\biggr) \ , }}
with corrections to both terms suppressed by $1/r^2$. Matching \smr\ at
small $r$ to \lgr\ at large $r$, one finds $\beta \ll \alpha$ and
\eqn\fiveeighteen{A = \frac{(r_+-r_-)^\ell\ \omega^{\ell+\half}
\ \Gamma (\ell+1) \Gamma (\ell+1+i\frac{\omega-m\Omega}{2\pi T_H})}{2^{\ell+
\half}\ \Gamma(\ell+\frac{3}{2}) \Gamma(2\ell+1) \Gamma
(1+i\frac{\omega - m\Omega}{2\pi T_H})}\ \alpha \equiv N\alpha  \ . } 

\subsec{Absorption}

The conserved flux associated to the radial wave equation \weq\ is 
\eqn\fivenineteen{f= \frac{2\pi}{i} (R^* \Delta \partial_r R-R\Delta
\partial_r R^*)\ .}
Using \bas\ the incoming flux from infinity is approximately
\eqn\fivetwenty{f_{\rm in} = 2 |\alpha|^2\ .}
The flux across the horizon follows from \hfn
\eqn\fivetwentyone{f_{\rm abs} = \frac{(\omega - m\Omega)}{T_H} (r_+ - r_-)
|N|^2 | \alpha |^2  \ .}
The absorption cross section of the radial problem is
\eqn\fivetwentythree{\sigma ^{\ell,m} 
 = \frac{f_{\rm abs}}{f_{\rm in}}
 =  { (\omega - m\Omega) {\cal A} \over 2 }  |N|^2\ .
}
where we used $(r_+ - r_-)  = {\cal A} T_H $. In order to convert the
partial wave cross sections to the usual plane wave cross sections
we have to multiply \fivetwentythree\ by $\pi/\omega^2$. 
For $\ell=0$ we find
\eqn\fivetwentyfour{\sigma^0_{\rm abs} = {\cal A}
\ , }
In fact for all black holes the low energy absorption cross section
is proportional to the area of the horizon \gibbons . 
For $\ell=1$ we find
\eqn\fivetwentyfive{\sigma^{1,m}_{\rm abs} = 
\frac{\omega {\cal A}^3}{36} \left(T^2_H + \frac{(\omega -m
 \Omega)^2}{4\pi^2 } 
 \right) (\omega-m\Omega) 
 \ .}
Note that the absorption cross section is negative for $\omega< m\Omega$
corresponding to superradiance. This means that the amplitude of the
reflected  wave is greater than that of the incident wave, and the
scattered wave mines energy from the black hole.

The decay rates are related to the scattering amplitudes by
\eqn\fivetwentysix{\Gamma^\ell = \frac{\sigma^{\ell,m}_{\rm
abs}}{e^{\frac{\omega-m\Omega}{T_H}} - 1}
 =  
{ \pi \Gamma(\ell +1)^2  \omega^{2\ell -1} T_H^{2\ell+1 }  {\cal A}^{2\ell+1}
\over 2^{2\ell +2} \Gamma(\ell +3/2)^2 
\Gamma(2\ell+1)^2 }
e^{ - \frac{\omega-m\Omega}{2T_H} } 
|\Gamma( \ell+1 +i{\omega - m\Omega \over 2\pi T_H} )|^2 \ .}
Note that the decay rate, unlike the cross section, is positive for all
values of $\omega$. Note also that for a very non extremal black hole
the exponential factors simplify in the low energy region 
since $\omega \ll T_H \sim 1/M $. 
Near extremality, $T_H\to0$ and $\Gamma^\ell$ reduces to
\eqn\fivetwentyseven{\eqalign{\Gamma^\ell\to 0 & \quad {\rm for} \quad 
\omega > m\Omega\ ,\cr 
\Gamma^\ell 
\to | \sigma^\ell_{\rm abs} | & 
\quad {\rm for}  \quad \omega < m \Omega\ .}}
In particular the $\ell > 0$ decay rate does not vanish for $T_H=0$, and is
dominated by $\omega \ltwid m \Omega$ emissions.

\subsec{ The Near-BPS and Extremal Limits in Einstein-Maxwell Gravity}

A black hole with generic values of $a$, $M$, and $Q$ is quantum
mechanically unstable due to both Hawking and superradiant emissions.
The exceptional, stable case is achieved in the limit of BPS-saturation:
\eqn\bps{\eqalign{M & = Q\ ,\cr
a & = 0\ .}}
In the $N=2$ supersymmetric extension of Einstein-Maxwell gravity, 
\bps\ represents a supersymmetric BPS state of the black hole
\refs\gihu.\foot{Unlike the $d=5$ case \refs\bmpv, the
addition of angular momentum in $d=4$ always breaks supersymmetry
\refs\hml.}

When the angular momentum is included, a black hole can be extremal but
not BPS-saturated. Extremality occurs when the horizon is at double zero
of $\Delta$,
\eqn\fourtwo{r_+ = r_-\ .}
This implies
\eqn\fourthree{\eqalign{M^2 & = Q^2 + a^2\ ,\cr
T_H & = 0\ .}}
In the extremal limit there is no Hawking radiation. However, if $a>0$
the black hole still decays through superradiant emission.  The black
hole loses its angular momentum much 
 as a $Q=M$ black hole loses its
charge in a theory with (mass/charge) $<$ 1 particles: Real pair creation
of $\ell  \not= 0$ particles ocurrs in the Kerr-Newman
ergosphere.
Actually, in a theory containing particles 
with (mass/charge) $<$ 1, like the
real
world, both extremal limits look very similar.

We wish to analyze low-lying excitations of the black hole about the BPS
limit. To do so we expand in the excitation energy
\eqn\fourfour{ \E \equiv M-Q\ .}
taking $E/Q \ll 1 $ and $Q \gg 1$ in Planck units. 
Since $a^2$ is bounded from above by $M^2-Q^2$ (greater values give a
naked singularity), this implies that $a^2$ is of order $a^2 \sim \E
Q$. To leading order in $\E$, we then find
\eqn\thap{\eqalign{r_\pm &\simeq Q \pm \sqrt{2 \E Q - a^2}\ ,\cr
{\cal A} & \simeq 4 \pi (Q^2 + 2Q \sqrt{2\E Q - a^2})\ ,\cr
T_H & \simeq \frac{\sqrt{2\E Q-a^2}}{2\pi Q^2}\ ,\cr
\Omega & \simeq \frac{a}{Q^2}\ .}}
The black hole will emit Hawking quanta with frequencies of order $T_H$
and superradiant quanta with frequencies of order $\Omega$. In both
cases this gives
\eqn\wmq{\omega \sim \sqrt{\frac{\E}{Q^3}}\ .}
This implies that the greybody factors in Hawking emission will be 
correctly given by formula \fivetwentysix\ since the only assumptions
that went into that calculation were  that $ \omega \ll  1/M $
and $\Omega \ll 1/M$  which are
certainly obeyed by \wmq \thap .
%furthermore, the eigenvalues of the angular Laplacian in \sae \hldef\  are
%approximated by
%
%\eqn\laap{\lambda \simeq A \simeq \ell (\ell + 1) \ .}
%

%\newsec{Fermions}

%In this section we consider emission and absorption of massless
%spacetime fermions obeying the Dirac equation $\not\kern-0.4em\nabla
%\Psi=0$ in the Kerr-Newman geometry. This equation was separated in
%\pgf. The radial wave equation is
%
%\eqn\sixone{\sqrt{\Delta} 
%\ \partial_r \sqrt{\Delta}\ \partial_r R + \frac{K^2 + i (r-M)K}{\Delta}
%\ R - 2i\omega r\  R - \lambda\ R=0}
%
%with $\Delta$, $\lambda$, and $K$ defined as in \ddef\ and \hldef. The
%separation constant $A$ here is given by the angular eigenvalue
%equation,
%
%\eqn\sixtwo{\frac{1}{\sin\theta}  \partial_\theta \sin \theta 
%\partial_\theta S
%+ \left[\left(\half + a\omega \cos \theta\right)^2 -
%\left(\frac{m-\half\cos\theta}{\sin\theta}\right)^2 - \half + 2a\omega m -
%a^2\omega^2\right] S=-AS\ .}
%
%For small $a, \omega$,
%
%\eqn\sixthree{\lambda\simeq A \simeq \ell(\ell + 1) + \frac{3}{4} \quad
%, \quad \ell E z+\half}
%
%with half-odd integer $\ell$.

%The far region wave equation is
%
%\eqn\sixfour{\frac{1}{r} \partial_r r \partial_r R + \omega^2 R -
%\frac{i\omega}{r} R - \frac{\lambda}{r^2} R=0}
%

\subsec{ Semiclassical Derivation of the Effective Low Energy Theory}
% Low energy thermodynamics and conformal field theory}

%There have been suggestions in the literature that 
%the black hole entropy and low energy dynamics is
%related to that of a highly excited fundamental
%string \lenny \ssc . Here we start from the black hole
%solution and we will infer, from general principles, the
%form of the low energy field theory describing the system.
To begin with, the excess energy $\E \equiv M-Q$ of a near-BPS,
non-rotating black hole is related to the Hawking temperature by
\eqn\etr{\E = 2\pi^2 Q^3 T^2_H\ .}
In $D$ spacetime dimensions the energy of a weakly-interacting field
theory scales like $T^D$. Hence \etr\ indicates that the quantum states
of the black hole are described by a $D=2$ field theory.  Since we are
taking $\E $ to be much less than any other mass scale in the
problem,\foot{ Except the black hole mass gap, see below.} (in particular
$\frac{1}{Q}$) this should be a $D=2$ conformal field theory, with
associated central charge $c$. The exact
energy-temperature relation for a $D=2$ conformal field theory is 
\eqn\twt{\E = \frac{\pi}{12} L\, c\, T^2_H\ ,}
where $L$ is the volume of the one-dimensional space. 
This relation is valid if $L$ is large compared to the typical
wavelengths
of the particles\foot{ Of course 
there are other situations, such as conformal field
theories with  a small mass gap due to twisted sectors, in which
$L$ is not large but  \etr\ remains valid.}
\eqn\largel{
L \gg { 1 \over T_H \ .} 
}
The energy 
 $\E $ is related
to $L_0$ by 
\eqn\seventhree{\E =\frac{2\pi}{L} L_0\ .}
Comparing \twt\ and \etr\ we learn that 
\eqn\qlc{c\, L = 24\pi Q^3\ .}
The group $SU(2)$ of global space rotations is a symmetry of the Hilbert
space of black hole states.  The Noether currents of this symmetry
generate an $SU(2)$ Kac-Moody algebra within the conformal field theory.
Let $j$ denote the generator of the $U(1)$ subgroup of rotations about
the $z$-axis.  $j$ obeys
\eqn\sevenfive{[j_m\ ,\ j_n] = \frac{km}{2}\ \delta_{m+n, 0}\ ,}
where the integer $k$ is the Kac-Moody level, and we have chosen the
normalization so that $j_0$ has half integral eigenvalues.  The eigenvalues
of $j_0$ are  the $z$-components, $J_z$,  of angular momentum.  Standard
arguments using the Sugawara construction of the stress
tensor imply that $L_0$ is bounded according to
\eqn\ibd{L_0 \geq \frac{J^2_z}{k}\ .}
This bound is saturated by the extremal states 
%$ e^{iJ_z\phi}| 0\rangle $,
%
\eqn\exb{e^{iJ_z\phi}| 0 \rangle \,}
where the $U(1)$ boson $\phi\sim\phi+2\pi$ is the angle about the
$z$-axis defined by $j={k \over 2 } \partial\phi$.
A similar bound can be derived in a completely different manner from the
condition $r_+ \geq r_-$;  i.e., the absence of a naked singularity
in the spacetime solutions.  Using \thap\ this implies 
\eqn\sevennine{2\E  Q \geq a^2\ .}
which, together with  \qlc , yields
\eqn\ljc{L_0 \geq \frac{6J^2_z}{c}\ .}
Consistency of the $2D$ effective field theory with the spacetime
analysis requires that \ibd\ and \ljc\ are the same bound. This yields
\eqn\kcv{c=6\ k\ .}
This is exactly the relation between $k$ and $c$ encountered in a chiral
$(0,4)$ superalgebra. This result is especially striking in that
so far we have been discussing the purely bosonic Einstein-Maxwell theory
and have not assumed or used supersymmetry in any way!

In a supersymmetric theory the black hole ground state is a soliton
preserving four supersymmetries.\foot{As is the case for extreme
Reisner-Nordstrom solutions of N=2, N=4 and N=8 supergravity.} 
We are considering here a chiral theory with only right movers,
so we must have a $(0,4)$ supersymmetry algebra. This algebra contains 
an $SU(2)_R$ symmetry, which in fact is the same as the 
$SU(2)$ symmetry of spatial rotations considered above.
This implies that $c=6 k$.
The black hole mass gap is then the energy of the lowest lying
 excitation. This is the extremal state \exb\ with
$J_z=\half$. Using \ibd, \kcv\ and \qlc; it has energy
\eqn\seventhirteen{\E_{\rm gap} = \frac{1}{8 Q^3}\ .}

The existence and value of the black hole mass gap was first derived
using fundamental string theory in \mss . Here we see it can be derived in a
purely semiclassical analysis, without any reference to fundamental
string theory. 

\subsec{Microscopic Decay Rates}

 We have seen that the effective string picture correctly 
reproduces the thermodynamic behavior of a near-BPS Kerr-Newman 
black hole. In section 2.9 we semiclassically computed 
the rate at which such a macroscopic black hole emits scalars as 
a function of emitted energy and angular momentum. 
In the effective string picture such decays arise from 
a coupling of the spacetime scalar $\phi$ to an operator 
$\cal O$ in the conformal field theory. These are 
of the general form 
\eqn\ifts{ S_{int} \sim \int dt d\s {\partial}^n \phi (0,t)
  {\cal O}( \s +t).}
The spatial argument of $\phi$ is $0$ because we  have 
taken the black hole to be at $x=0$. $\sigma$ here is 
the spatial coordinate in the conformal field theory. 
$\cal O$ depends only on $\sigma +t$ because it is chiral theory. 

We know of no principle, without recourse to string theory,
which allows us to determine the numerical 
coefficients in front of the 
couplings in \ifts . However we will be able to determine the 
energy and angular momentum dependence of the decay rates in 
great detail. A striking agreement between the macroscopic and 
microscopic decay rates will be found.

The amplitude $\cal M$ for the 
emission of a particle with energy $\omega$ 
has an internal contribution
\def\sp{\sigma^+}
\eqn\mint{{\cal M} \sim \int d\sp 
 \langle f | {\cal O}(\sp) | i \rangle 
e^{-i\omega \sp },}
where $\sp \equiv \sigma+t$. Squaring and summing over final states 
 gives
\eqn\ampltfinal{
 \sum_{f} | {\cal M} |^2 \sim 
  \int d\sp d{\sp}^\prime  \langle
 i  | {\cal O}^\dagger (\sp)
 {\cal O} ({\sp}^\prime )  | i \rangle 
 e^{- i \omega (\sp-{\sp}^\prime) }
}
for the rate, 
where we have used the fact that 
the sum over final states produces an identity matrix.
To compare with the macroscopic decay rate we should 
thermally average, at temperature $T_H$ with an angular 
potential $\Omega$, over the initial state. This  potential 
implies that a state with $U(1)$ charge $m$ is 
weigted by $exp(-(\omega-m\Omega)/T_H)$.
The rate is then proportional to the thermal correlation function
\eqn\thrmal{
\int d\sp 
\langle  {\cal O}^\dagger (0)
 {\cal O} (\sp)  \rangle _{T_H
} 
e^{-i (\omega-m\Omega) \sp}
}
$m$ arises here as the $U(1)$ charge of ${\cal O}$. 
The shift in $\omega$ implements the effects of the angular potential.

This correlation function is completely determined by the 
conformal weight of the operator ${\cal O}$.
If the operator $\cal O $ has conformal weight $\Delta$ 
then the
correlation function becomes 
\eqn\corr{
\langle {\cal O}^\dagger (0)
 {\cal O} (\sp) \rangle _{T_H}
 \sim   \left[{   \pi T_H  \over  \sinh(  \pi T_H\sp  )}
\right]^{ 2 \Delta }.   
}
Note that according to \largel\ we can ignore the periodicity along
the spatial direction, and take it to be infinite when we calculate 
\corr . The periodicity along euclidean time translates into
\eqn\perid{
\sp \sim \sp + i 2/T_H .}
Hence we must evaluate the integral 
\eqn\intg{
\int d\sp e^{-i(\omega-m\Omega) \sp } 
\left[ {  T_H  \over (\sinh(  \pi T_H  \sp
))}
\right]^{2
\Delta}.}
This Fourier transform should be performed with an $i\epsilon$
prescription for dealing with the pole at $x=0$. The two different 
prescriptions correspond to absorption or emission.
The
one corresponding to emission gives
\eqn\integral{
 \int d\sp 
 e^{-i(\omega-m\Omega) (\sp-i\epsilon)}  
\left[ {  T_H  \over (\sinh(   \pi T_H \sp  
))}
\right]^{2
\Delta} \sim 
% e^{ - (\omega-m\Omega) /2T_H } 
%\int d\sp 
% e^{-i(\omega-m\Omega) \sp }   \left[ {  T_H  \over (cosh(  \pi T_H \sp
%))}
%\right]^{2
%\Delta}=
 (T_H)^{ 2 \Delta -1} e^{ - { \omega-m\Omega\over  2T_H} }
  \left| \Gamma( \Delta + i { (\omega-m\Omega)  \over
2 \pi T_H } )\right|^2
% \Gamma( \Delta - i { (\omega-m\Omega) \over2 \pi T_H } )
}

The next problem is to determine $\Delta$.  The coupling 
\ifts\ and the allowed operators $\cal O$ 
are restricted by the symmetries of the theory. 
%If we imagine a very low energy coupling, or the limit of zero energy,
%the coupling to $\cal O$ is a supersymmetric deformation of the 
%theory.
 The simplest way of satisfying this restrictions is when
the  integral of $\cal O$ is invariant under 
supersymmetry transformations of the conformal field theory
and is single-valued on the circle ($i.e.$ it is not a 
twist field)\foot{
Though this gives the desired answer, the justification
of this second assumption is unclear since the states
\exb\ themselves are not single valued.}. Under these conditions there is a bound 
relating the conformal weight of the operator and the U(1) charge
\eqn\bnd{\Delta \ge \ell +1. }
Operators which saturate the bound are of the form 
\eqn\sat{{\cal O}=\{ G_{1/2}, {\cal C} \},}
with $\cal C$ a chiral primary and $G$ a supercharge. 
The leading contribution in the 
low energy expansion comes from the 
lowest value of $\Delta$, so we conclude   
\eqn\bnd{\Delta = \ell +1. }

An additional contribution to the energy
dependence  arises from the external spacetime part of
the interaction in \ifts . For a mode of angular momentum $\ell$,
the first $\ell -1$ spatial derivatives of $\phi$ vanish at the origin. 
Hence $n$ must be at least $\ell$ in \ifts . The leading contribution
is for $n=\ell$. Since the field $\phi$ is massless derivatives 
give powers of $\omega$.  In other words, we need at least $\ell$ powers 
of the spatial momentum to match the SO(3) transformation properties
of ${\cal O}$. Hence, there is an additional power of 
$\omega^{2\ell}$ in the rate. Multiplying by a factor of 
$1/\omega$ for the normalization of the outgoing state and 
restoring powers of $Q$ with dimensional analysis, the microscopic
rate is 
\eqn\drt{\Gamma^{\ell}
\sim \omega^{ 2 \ell-1 }Q^{4\ell +2 }(T_H)^{ 2 \ell +1}
 e^{ -{ \omega-m\Omega \over 2 \pi T_H} }
 \left|  \Gamma( \ell +1 + i { (\omega-m\Omega)  \over
2 \pi T_H } ) \right|^2
% \Gamma( \ell +1 - i { (\omega-m\Omega) \over2 \pi T_H } )
}
in agreement with \fivetwentysix\ noting that ${\cal A} \sim Q^2$. 
%In \ssc\ it was shown that the low energy absorption 
%cross section for scalars 
%can be reproduced for all black holes by a $c=6$ effective string

\newsec{ Five-Dimensional Black Holes}

In this section we consider the interactions of five-dimensional, 
non-rotating black holes 
in string theory with a massless scalar, extending 
the results of \dm \jmas \gk \cgkt\ to include orbital angular
 momentum for the scalar field.
Qualitatively new features arise in five dimensions
because the spatial rotation group is $SU(2)\times SU(2)$, 
and one $SU(2)$ is carried by 
right-movers while the other is carried by  
left-movers \bmpv\ on the effective string. A single left-moving 
and a single right-moving fermion can collide and create 
an outgoing boson. The rate for this 
process will involve a left and a right $\it fermionic$ 
thermal factor. We will find that
these factors arise in the greybody calculation. It is 
fascinating that one can `see' that the black hole is in 
part made of fermions in such a purely bosonic calculation!

\subsec{ Semiclassical Scattering }

We consider the scattering of scalars from a five dimensional 
black hole in the dilute gas approximation considered in
\jmas . We follow the notation of \jmas, where further details of 
the geometry may also 
be found. The wave equation is
\eqn\waveeqn{
{g \over r^3} {d \over dr } r^3 g { d \phi \over dr }
+ { g \over r^2 } \nabla^2_\theta \phi  +
\omega^2 f \phi =0,
}
where 
\eqn\fandg{
f = (1 + {r_1^2 \over r^2 })
(1 + {r_5^2 \over r^2 })(1 + {r_n^2 \over r^2 }),~~~~~~~~~~~~~
g =1 - {r_0^2 \over r^2 }. 
}
$\nabla^2_\theta$ is the angular Laplacian which has eigenvalues
${\ell}({\ell}+2)$ in five dimensions and ${\ell}$ is an integer
 which labels the
orbital angular momentum. The rotation group SO(4) can be decomposed
as SO(4)  $\sim$ SU(2)$_L\times$SU(2)$_R$. 
The representations appearing here corresponds to
$({\ell}/2,{\ell}/2)$ 
under the
two SU(2)'s. The degeneracy of this representation is $({\ell}+1)^2$
corresponding
to the different allowed $J_z$ values of each SU(2).

We consider low energies satisfying $\omega \ll 1/r_1, 1/r_5 $ and 
we 
 divide space into a far region $r \gg r_1,r_5 $ and a near region
$r \ll  1/\omega$ and we will match the solutions in the overlapping
region.
In the far region we write $ \phi = {1\over r} \psi $ and the 
equation for $\psi$ becomes, with $\rho = \omega r$,
\eqn\fareqn{
{d^2 \psi \over d\rho^2 } + {1\over \rho } { d \psi \over d\rho} +
\left( 1 -{  ({\ell}+1)^2  \over \rho^2 }
\right) \psi =0
}
The solutions are the Bessel functions 
%where $\tilde r^2 = r^2_1 + r^2_5 + r^2_n \sim r_1^2 + r_5^2 $.
%Defining the parameter $\nu^2 =({\ell}+1)^2 - \omega^2 \tilde r^2 $
%the solution is
\eqn\solfar{
\phi = { 1 \over \rho } \left[ \alpha J_{\ell +1} (\rho) + \beta 
J_{-\ell -1 }(\rho) \right].
}
   From the large $\rho$ behavior the incoming
flux is found to be
\eqn\flux{f_{in} = Im( \phi^* r^3 \partial_r \phi )
= { 1 \over  \pi\omega^2 } | \alpha e^{i (\ell +1) \pi/2} + 
\beta e^{-i (\ell +1) \pi/2} |^2.
}}
On the other hand the small $\rho$ behavior of the far region 
solution
is 
\eqn\smallrho{
\phi = { 1 \over \rho } \left[
\alpha ({\rho\over 2})^{\ell +1}
( { 1 \over \Gamma(\ell +2 ) } - {\cal O}(\rho^2) ) +
\beta ({\rho\over 2})^{-\ell -1} 
( { 1 \over \Gamma(-\ell) } - {\cal O}(\rho^2) ) \right].
}
Since \smallrho\ has a pole for integer  $\ell $ it is convenient
to keep $\ell $ near an integer value during the calculation and make
it integer at the end. 
Now we turn to the solution in the near region $r \ll 1/\omega $. 
Defining $v = r_0^2/r^2$, the near region wave equation is
\eqn\nearone{
(1-v)^2 {d^2 \phi \over dv^2 } - (1-v) {d \phi \over dv }
+\left( C + {D \over v} + { E \over v^2 } \right) \phi =0
}
where 
\eqn\cdef{
C = \left( \omega r_n r_1 r_5 \over 2 r_0^2 \right)^2 ~,~~~~~~
D =  { \omega ^2  r_1^2 r_5^2 \over 4 r_0^2 }    
 + { \ell ({\ell}+2)\over 4} ~,~~~~~
E =  -{\ell ({\ell}+2)\over 4}.
}
Defining 
\eqn\defoff{
\phi = v^{- \ell/2} (1-v)^{-i {\omega \over 4 \pi T_H}} A  F }
with $A$ a constant, 
we find that the solution to \nearone\ with only ingoing flux
at the horizon is given by \defoff\ with
 $F = F(\alpha, \beta, \gamma, 1-v )$, 
a  hypergeometric function,  with
%\eqn\defpq{
%p =\ell /2~~~~~~~~~~~~~~~q^2 = -(C+D+E) =
% - \omega^2 r_0^2(
%\cosh^2  \alpha \cosh^2\gamma \cosh^2\sigma -1 )/4 \sim 
%( { - i \over 4 \pi } { \omega \over T_H } )^2
%}
%we obtain a hypergeometric equation for $F$.
%Demanding that we only have an incoming wave at the horizon fixes
%$\phi$ to be of the form \defoff\ with $q$ as in \defpq . 
%The solution is
%\eqn\solution{
%\phi = v^p (1-v)^q A F(\alpha, \beta, \gamma, 1-v ),
%}
%with $F$ the hypergeometric function, $A$ an undetermined
%constant, and
\eqn\angles{
\gamma = 1 + 2q,~~~~~ 
\alpha =- \ell/2 + q + i \sqrt{C},~~~~~~~
\beta = -\ell/2 + q - i \sqrt{C},~~~~~q = i {\omega \over 4 \pi T_H }
}
 The behavior for small $v$ is 
\eqn\smallv{\eqalign{
\phi \sim & A v^{-\ell/2} \left\{
{ \Gamma(1+2q) \Gamma(1+ \ell ) \over \Gamma( 1+\ell/2 +q-i\sqrt{C} )
\Gamma( 1+ \ell/2 + q+i\sqrt{C} ) } ( 1 + {\cal O}(v) )  
\right. \cr +& \left. 
v^{1 + \ell}{ \Gamma(1+2q) \Gamma(-1 -\ell) \over 
\Gamma( -\ell/2+ q +i\sqrt{C} )
\Gamma(-\ell/2+ q-i\sqrt{C} ) } ( 1 + {\cal O}(v) ) \right\}.
}}

Matching solutions in the overlapping region 
% at $r = r_m$
and anticipating that $\beta \ll \alpha $ in 
\solfar , we find
\eqn\defalpha{
\alpha/2 = A ( \omega r_0/2)^{-\ell} \Gamma(1+\ell) \Gamma(2 + \ell)
\left[ { \Gamma(1 + 2 q) \over \Gamma( 1 + \ell/2 + q -i\sqrt{C} )
\Gamma( 1 + \ell/2 + q +i\sqrt{C} )} \right].
}
The absorbed flux is 
\eqn\absflux{
f_{abs} = Im( \phi^* g r^3 \p_r \phi) = 2 r_0^2 
Im( \phi^* g \p_v \phi ) = 4  |q| r_0^2 |A|^2.
}
Hence the absorption cross section for the radial problem is
given by the ratio  of the two fluxes \absflux \flux . The plane
wave cross section is obtained multiplying by $4\pi/\omega^3$ and
we obtain
\eqn\crossradnew{
\sigma^\ell_{\rm abs} = 
%{2 \pi^2 } r_1r_5 r_0 \cosh\sigma 
{\cal A}_H (r_0 \omega )^{2 \ell} 
\left| {2^\ell \over \Gamma({\ell}+1)
\Gamma({\ell}+2 )}\right |^2 
  \left| {\Gamma( { ({\ell}+2) \over 2 } + i { \omega\over 4 \pi T_L} )
\Gamma(  { ({\ell}+2) \over 2 }+ i{\omega\over4\pi T_R}  )
\over \Gamma(1+ i {\omega\over2\pi T_H})  } \right |^2 
}
We see the same left-right structure that  was  found 
for ${\ell}=0$ in \jmas, but  with the difference 
that the factor of ${\ell}$ appears
inside the gamma functions. To evaluate this 
we need the absolute value of 
$\Gamma (n/2 + ib) $.  Properties of
gamma functions can be used to reduce  $\Gamma (n/2 + ib) 
\Gamma (n/2 -ib)$ to
the form $\Gamma(ib) \Gamma(1-ib)$ or $\Gamma(1/2 +ib) \Gamma(1/2 -ib)$ 
(times a polynomial in $b$). 
While $\Gamma(ib) \Gamma(1-ib)$ is proportional to $b/\sinh \pi b $,   
$\Gamma(1/2 +ib) \Gamma(1/2 -ib)$ is proportional to 
$1/\cosh \pi b $. This $\cosh$ will eventually translate
into a fermionic as opposed to bosonic ocupation factor, and
this happens for odd ${\ell}$ as it should. 
For example, in the case of ${\ell}=1$ 
\eqn\lone{
\sigma^1_{\rm abs} = 
 {\pi^3 \over 8} (r_1 r_5 )^4 \omega 
[ ( 2 \pi T_L)^2 + \omega^2 ] [ (2 \pi T_R )^2 + \omega^2 ] 
{ e^{\omega/T_H } -1 \over ( e^{ \omega \over 2 T_R} + 1 )
( e^{ \omega \over 2 T_L} + 1 ).
}
}
The corresponding decay rate seems
to come from two particles colliding, where both particles are 
fermions,as one expects from the decomposition of ${\ell}=1$ as
$(1/2,1/2)$ under the two SU(2)'s! We shall see that 
this coincides beautifully 
with the effective string picture. 
In general fermions appear for odd ${\ell}$ and bosons for even ${\ell}$.

The Hawking rate for emitting particles with
angular momentum  is obtained by multiplying \crossradnew\ by the usual 
Hawking thermal
factor producing 
%%%%% 
\eqn\hawrate{
\Gamma_H = { 2^{4\ell +4}
 \pi^{2l +3}
(r_1^2 r_5^2 T_L T_R)^{{\ell}+1}  \omega^{2l-1}
 \over |\Gamma({\ell}+1) \Gamma({\ell}+2)|^2 }
e^{-\omega \over 2 T_H} | \Gamma(1+ {\ell \over 2}+ i{\omega \over
4 \pi T_L} ) \Gamma(1+ {\ell\over 2}+ i{\omega \over
4 \pi T_R} )|^2 \ .
}
%We note for future comparison that the factor 
%$(r_1^2 r_5^2)^{{\ell}+1} \sim g^{2{\ell}+2} (Q_1
%Q_5)^{{\ell}+1}
%$.

\subsec{Microscopic Decay Rate}

It would be of interest to see how much of the structure of the 
effective string 
for the five-dimensional black hole can be deduced, as 
in the four-dimensional case of the previous sections, 
without fundamental string theory. 
However in this section our main goal is to 
understand the emission of angular momentum and
 we shall simply assume  the 
structure 
implied by string theory. String theory says that the low energy
dynamics is described by a two dimensional conformal field 
theory \ascv\ which is  good description of the low energy 
dynamics in the black hole region \jmmlow . Some twisted sectors
of this SCFT contain excitations which can be viewed as the excitations
of a long multiply wound string.
 Some things are more clear 
in this picture of a multiply wound string and some things are
more clear thinking about  the full conformal field theory, so it
is useful to keep both in mind.

In section 2.12 conformal field theoretic arguments were given for 
determining the microscopic decay rate. The same arguments 
are applicable here, with a few modifications as follows. 
 Since there are 
right movers as well as left movers, 
the rate involves the product of two factors of the 
form \integral, one at temperature $T_L$ 
and conformal weight $\Delta_L$ and one at 
temperature $T_R$ and conformal weight $\Delta_R$.
The bounds on $\Delta$ give $\Delta_L=\Delta_R=1+\ell/2$. 
%Operators with $\Delta_L-\Delta_R = \pm 1/2 $  couple
%to spacetime fermions. 
In this section we are considering $\Omega=0$ black 
holes so there is no shift of $\omega$, but it is easy to see
how to extend the calculation for general $\Omega$.

It is possible to give the general form of the operator 
with  
the correct weights. 
For a scalar arising from metric deformations of the internal $T^4$
(an internally polarized graviton), 
the  operator 
is of the form
\eqn\coupling{
\epsilon_{IJ} \omega^\ell {\cal O}^I_{{\ell}/2,L} {\cal O}^J_{{\ell}/2,R} 
}
where $\epsilon_{IJ}$ indicates the polarization of the graviton
in the internal directions, and 
the factors of $\omega$ arise as before because the spacetime field 
must be acted on by  ${\ell}$ derivatives. 
 The operator ${\cal O}_{L}{\cal O}_R$ involves the fields propagating on the 
effective string and has angular 
momentum $({\ell}/2,{\ell}/2)$ under the SU(2)$_L \times$SU(2)$_R \sim $  SO(4)
symmetry. It must also carry the indices $IJ$ in order to be contracted with
the graviton polarization tensor. The $I$ index is carried by the 
bosonic field $\partial X^I$
living on the brane. 
The simplest such operators $\cal O$ are of the form
\eqn\operators{
{\cal O}^I_{{\ell}/2} \sim \partial X^I \psi_{i_1} \psi_{i_2} \cdots 
\psi_{i_\ell},
}
where $\psi_i $ are different families of fermions propagating on the 
string. 
The operator \operators\ 
has $\Delta=1+{\ell}/2$, precisely as needed for agreement with the factors 
in \hawrate.

Putting this all together, our final result for the 
Hawking emission rate is, up to numerical coefficients, 
\eqn\rate{
\Gamma \sim  (  g^2 Q_1 Q_5 T_L T_R )^{{\ell}+1}
\omega^{2\ell -1} e^{-{\omega \over 2 T_H}}
|\Gamma( {\ell}/2 +1 + i {\omega \over 4 \pi T_L })|^2
|\Gamma( {\ell}/2 +1 + i {\omega \over 4 \pi T_R })|^2.
}
In this expression we have multiplied by a factor of $1/\omega $ 
from the normalization of  the outgoing scalar and a factor of
$\omega^{2\ell}$ 
from squaring the  $\omega$ factors in the vertex operator \coupling . 
The factor of $g^{2{\ell}+2}$ comes from the
fact that in string theory this is a disc amplitude with
a closed string plus $2 {\ell}+2 $ open strings (${\ell}+1$ right moving and
${\ell}+1$
left moving). Finally the factor of $ (Q_1Q_5)^{{\ell}+1}$ can be
understood
from the fact that one can create this many different families 
of open strings.

Once we remember that $g^2Q_1Q_5 = r_1^2r_5^2$ we see that 
\rate\ agrees precisely with \hawrate. Once again we find 
detailed agreement between the macroscopic and microscopic 
descriptions of black hole dynamics. It would be interesting 
to calculate, as it was done for the $l=0$ case \dm\ 
and the fixed scalar case \cgkt , the 
precise numerical coefficient in \rate\ and compare it with
\hawrate . As an aside,  note that the full energy dependence
of the cross sections for the fixed scalars \cgkt\  comes from
the fact that it couples to an operator with conformal
weights $\Delta_L =\Delta_R =2$.

{\bf Acknowledgments}

We  thank I. Klebanov, S. Mathur, L. Susskind and C. Vafa 
for illuminating discussions. We also thank S. Gubser for
pointing out a numerical error and for informing us that
he had done some similar calculations \steve .
 
The research of J.M. was supported in part by DOE grant 
DE-FG02-96ER40559.

\listrefs

\bye